\documentclass[10pt,conference]{IEEEtran}
\IEEEoverridecommandlockouts

\usepackage{cite}
\usepackage{amsmath,amssymb,amsfonts}
\usepackage{algorithmic}
\usepackage{graphicx}
\usepackage{textcomp}
\usepackage[dvipsnames]{xcolor}
\usepackage{xcolor}
\usepackage{tabularx}
\usepackage{booktabs}
\usepackage{hyperref} 
\usepackage{comment}
\usepackage{float}
\usepackage{adjustbox}
\usepackage{pgf-pie}  
\usepackage{pgfplots} 
\pgfplotsset{compat=1.8} 
\usepackage{tikz}
\usepackage{colortbl}
\usepackage{soul} %
\usepackage[para,online,flushleft]{threeparttable}

\usepackage{multirow}
\newcommand{\ra}[1]{\renewcommand{\arraystretch}{#1}}

\usepackage[strict]{changepage} %
\usepackage{framed} %

\newcommand{\rqone}{To what extent do SE participant studies report the diversity of participants?}
\newcommand{\rqtwo}{What  diversity categories are reported in SE participant studies?}
\newcommand{\rqthree}{What is the function of diversity in SE participant studies?}

\definecolor{formalshade}{rgb}{0.95,0.95,1}
\definecolor{darkblue}{rgb}{0.0, 0.0, 0.55}
\newenvironment{quotebox}{%
    \vspace{-.1cm}
	\MakeFramed{\advance\hsize-\width\FrameRestore}%
	\noindent\hspace{-4.55pt}%
	\begin{adjustwidth}{}{7pt}%
		\vspace{-1pt}\vspace{1pt}%
	}
	{%
		\vspace{1pt}\end{adjustwidth}\endMakeFramed%
}

\sethlcolor{gray!15}

\def\BibTeX{{\rm B\kern-.05em{\sc i\kern-.025em b}\kern-.08em
    T\kern-.1667em\lower.7ex\hbox{E}\kern-.125emX}}

\title{Diversity Awareness in Software Engineering Participant Research}

\author{\IEEEauthorblockN{Riya Dutta}
\IEEEauthorblockA{
    \textit{Department of Computer Science and} \\
    \textit{Software Engineering} \\
    \textit{Concordia Univeristy} \\
    Montreal, Canada \\
    r\_utta@live.concordia.ca
}%
\\
\IEEEauthorblockN{Emad Shihab}
\IEEEauthorblockA{
    \textit{Department of Computer Science and} \\ 
    \textit{Software Engineering} \\
    \textit{Concordia Univeristy} \\
    Montreal, Canada \\
    emad.shihab@concordia.ca
}%
\and
\IEEEauthorblockN{Diego Elias Costa}
\IEEEauthorblockA{
    \textit{Department of Computer Science} \\ 
    \textit{Université du Québec à Montréal}  \\
    LATECE Lab \\
    Montreal, Canada \\
    costa.diego@uqam.ca
}
\\
\IEEEauthorblockN{Tanja Tajmel}
\IEEEauthorblockA{
    \textit{Centre for Engineering in Society} \\ 
    \textit{Concordia Univeristy}}
    Montreal, Canada \\
    tanja.tajmel@concordia.ca
}

\begin{document}
\maketitle

\begin{abstract}
Diversity and inclusion are necessary prerequisites for shaping technological innovation that benefits society as a whole. A common indicator of diversity consideration is the representation of different social groups among software engineering (SE) researchers, developers, and students. However, this does not necessarily entail that diversity is considered in the SE research itself. 

In our study, we examine how diversity is embedded in SE research, particularly research that involves participant studies. To this end, we have selected 79 research papers containing 105 participant studies spanning three years of  ICSE technical tracks. 
Using a content analytical approach, we identified how SE researchers report the various diversity categories of their study participants and investigated: 1) the extent to which participants are described, 2) what diversity categories are commonly reported, and 3) the function diversity serves in the SE studies.

We identified 12 different diversity categories reported in SE participant studies. Our results demonstrate that even though most SE studies report on the diversity of participants, SE research often emphasizes professional diversity data, such as occupation and work experience, over social diversity data, such as gender or location of the participants. 
Furthermore, our results show that participant diversity is seldom analyzed or reflected upon when SE researchers discuss their study results, outcome or limitations.

To help researchers self-assess their study diversity awareness, we propose a diversity awareness model and guidelines that SE researchers can apply to their research. With this study, we hope to shed light on a new approach to tackling the diversity and inclusion crisis in the SE field.

\end{abstract}


\section*{General Abstract}

Incorporating diversity considerations in research, development, and innovation has become an increasingly important topic. It is a well-known fact that diverse teams produce better outcomes, whereas the lack of diversity might result in biased and discriminatory technologies. Therefore, the inclusion of diverse stakeholders is considered paramount for the creation of an ethical and social-responsible future. With this study, we aim to contribute to the conversation on how EDI (equity, diversity, inclusion) can be implemented in Software Engineering (SE) research. In our study, we focus on SE research that includes research participants since this is an evident opportunity to consider diversity, and we investigate to which extent and with what purpose SE researchers consider and report diversity in their research papers. Our results demonstrate that only a few studies do not consider diversity at all, however, the examined studies differ greatly in the range of the consideration and reporting of diversity. From these outcomes, we draw the conclusion of differences in the diversity awareness among SE researchers. Finally, we propose a model of diversity awareness for participant studies as a tool to support SE researchers in reflecting on diversity and incorporating it systematically in their research.

\section{Introduction}
\label{sec:introduction}

In recent years, the consideration of the diversity of people who are involved in research and innovation has become an increasingly important topic. Diversity of developers and researchers is regarded – and has already been proven - beneficial for numerous reasons: diverse teams produce better outcomes \cite{Menezes&Prikladnicki:2018, Pieterse&Kourie:2006,Patrick&Kumar:2012}, research and development that systematically includes diversity are of benefit for a broader population whereas the lack of diversity might result in biased and discriminatory technologies \cite{Schiebinger&Klinge:2011,Buolamwini&Gebru:2018,Tannerbaum&Ellis:2019}, the inclusion of diverse stakeholders is paramount for the creation of an ethical and social-responsible future \cite{Sarewitz:2005} \cite{VanOudheusden&Shelley-Egan:2021}, and, last but not least, excluding large parts of society from research and the development of future technologies violates human rights \cite{Oberleitner:2021,Al-Nashif:2021,PARL:2010}. One size does not fit all, and this applies to research and development, too. 
To mitigate the negative effects of non-diverse research and to stimulate researchers to incorporate diversity in their research, national and international research funding programs are developing policies and guidelines for equity, diversity, and inclusion (EDI) \cite{NSERC:2022}, for responsible research and innovation (RRI) \cite{EU:2013} to bind the research funding to increasingly important conditions: the sufficient consideration of EDI. 

In our study, we focus on research that includes research participants as we regard this as an evident opportunity for SE researchers to include diversity considerations in their research. Therefore, our research questions are as follows:
\begin{itemize}
    \item RQ1. \rqone 
    \item RQ2. \rqtwo
    \item RQ3. \rqthree
\end{itemize}

To capture all variants of diversity categories, we choose an inductive approach, which means: we do not predefine the categories to examine the content according to these categories. Instead, we reconstruct the diversity categories through open coding of the content. Therefore, for the purpose of this study, we choose a broader definition of diversity categories: A diversity category is a category based on which individuals (in our case, research participants) are distinguished and grouped. This open approach allows for identifying categories that might be relevant for research but are not included in the commonly discussed diversity categories such as gender, race, and ethnicity, amongst others \cite{Nkomo&Stewart:2006}.

Our results demonstrate that only a few studies do not consider diversity at all, however, the examined studies differ greatly in the range of the consideration and reporting of diversity. From these outcomes, we draw the conclusion of differences in diversity awareness among SE researchers. We propose a model of diversity awareness for participant studies as a tool to support SE researchers in reflecting on diversity and incorporating it systematically in their research.
Finally, we publish our classification coding scheme and the dataset used to conduct this study to encourage further studies on diversity awareness and facilitate replicability\footnote{\url{https://doi.org/10.5281/zenodo.7587076}}.

This paper is organized as follows. 
Section~\ref{sec:relatedWork} describes the related literature. 
We detail our study methodology in Section~\ref{sec:methodology} and present the results of our three research questions in Section~\ref{sec:results}.
Upon reflecting on the results, we discuss a model of diversity awareness and guidelines for the community in Section~\ref{sec:discussion}.
Finally, we discuss the limitations of our study in Section~\ref{sec:threats} and our final remarks in Section~\ref{sec:conclusion}.

\section{Related Work}
\label{sec:relatedWork}

Time and time again, diversity, and the importance of diversity, has been a topic of discussion for many researchers in various fields. This has led to different definitions of diversity emerging in different contexts \cite{Tamtik:2022, Zanoni&Janssens:2004, Nagappan&Zimmermann:2013}. In recent years, diversity, especially gender diversity, has received increasing attention in the field of SE \cite{Catolino&Palomba:2019, Burnett&Peters:2016, Padala&Mendez:2022, Bosu&Sultana:2019}. 

\noindent\textbf{Lack of Diversity in SE.}
The issue of lack of diversity, also referred to as the diversity crisis, in SE has been investigated by many SE research papers as can be seen in the IEEE special issue article by Albusays \textit{et al.} \cite{Albusays&Bjorn:2021}. The importance of diversity and inclusion in the field of SE has been clearly stated in the article with one of the questions being ``What are the relevant diversity parameters we should consider when exploring software development practices and technology'' \cite{Albusays&Bjorn:2021}. Tamtik~\cite{Tamtik:2022} notes that only understanding diversity in terms of gender, language, and socio-economic categories limits the understanding of diversity. Similarly, Rodriguez and Nadri~\textit{et al.} conclude that more perceived diversity aspects need to be considered in SE research \cite{Rodriguez&Nadri:2021}. One of the goals of our research is to address these challenges in the context of SE research, by (i) revealing how SE researchers are addressing diversity in their studies and reporting about it in their publications, and (ii) by providing SE researchers with guidelines they can use when conducting participant studies.  

Other work by Menezes and Prikladnicki~\cite{Menezes&Prikladnicki:2018} investigated various publications about diversity in SE by performing a literature review. They also investigated the impact of diversity on processes in software development by conducting semi-structured interviews. They concluded that many challenges still need to be overcome to make SE work environments more diverse. Our work complements the work by Menzes and Prikladnicki since we investigate to what extent SE research considers diversity, the types of categories used and the function of diversity.

\noindent\textbf{Human Values in SE Research.}
Perera~\textit{et al.} examine to what extent human values are considered in SE research instead of diversity \cite{Perera&Hussain:2020}. They use a similar methodology to ours, where they analyze research papers from top-tier SE conferences and journals to assess their consideration of human values. The found that very few SE publications consider human values.
Storey~\textit{et al.} conducted a similar study where they investigate the consideration of human aspects in SE research by analyzing SE research papers \cite{Storey&Ernst:2020}. They found most SE studies focus on technical aspects of SE even though these studies claim to impact human stakeholders. Storey~\textit{et al.} concluded that there is a need for SE studies to consider more human aspects and recommended a framework they created that can be used to consider more human and social aspects in SE research.

Although the work by Perera~\textit{et al.} \cite{Perera&Hussain:2020} and Storey \textit{et al.}~\cite{Storey&Ernst:2020} used a similar approach to ours, these two studies investigate the considerations of human aspects/values in SE research whereas we study the consideration of diversity in SE research, more specifically participant studies, while providing guidelines for SE researchers conducting participant studies.

The work closest to ours is the work by Lenarduzzi \textit{et al.}~\cite{Lenarduzzi&Dieste:2021} who investigated participant studies in SE, where they examined the current participant selection guidelines and practices in empirical SE research. They analyze existing guidelines for participant selection in SE and present the participant selection strategies being currently used in their results. Our work complements their study since we study actual research papers involving participants and create guidelines
for SE researchers, whereas their study only examined guidelines (and not actual research papers).

\section{Methodology}
\label{sec:methodology}

The goal of our research is to examine to which extent diversity is considered in SE research and to provide guidelines to SE researchers that they can use when conducting participant studies. As a methodological approach, we apply qualitative and quantitative content analysis \cite{krippendorff&Bock:2009} \cite{krippendorff:2018} where we used open coding and axial coding methods \cite{Saldana:2009} to examine software engineering research papers which include participant studies. Due to the novel nature of this study, we start by performing a pilot study to assess the viability of this study and develop an appropriate framework. 

\begin{figure*}
    \centering
    \includegraphics[width=\textwidth]{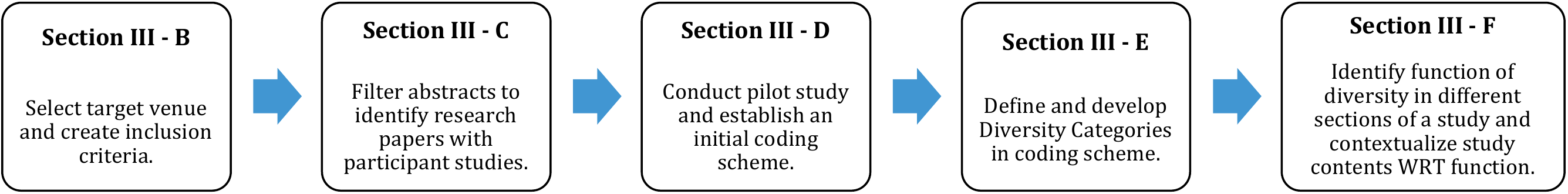}
    \caption{Overview of methodology as explained in Section \ref{sec:methodology}.}
    \label{fig:method-overview}
\end{figure*}

Fig.~\ref{fig:method-overview} shows an overview of our methodology. As the first step, we selected a venue to collect data for our study. Then, we conducted a pilot study to assess the viability of our study and establish a coding scheme. After the coding scheme was developed, we extended the data collection process to include more studies with participants. To collect our sample data, we filtered through abstracts of research papers from our selected venues to identify studies with participants. Once we gathered our data, the first author read each paper and classified each study according to our coding scheme. We leveraged a mixed methods approach that uses qualitative and quantitative methods to reach our results. 

\subsection{Research Questions}

To assess to what extent SE research papers with participants display diversity awareness, we first investigate if SE research papers report on diversity in their studies:
\\

\noindent \textbf{RQ1. \rqone}\\

\noindent After examining whether diversity is even reported, we set out to identify the kinds of diversity reported:\\

\noindent \textbf{RQ2. What  diversity categories are reported in SE participant studies?}\\

Finally, after assessing what kind of diversity categories can be identified in SE research with participants, we investigate if the reported diversity is actually analysed/leveraged throughout the research paper:\\

\noindent \textbf{RQ3. What is the function of diversity in SE participant studies?} \\

\subsection{Selecting the Target Venue}

To take a more systematic approach, we collected the sample for this research from the International Conference of Software Engineering proceedings (ICSE). We choose ICSE as our venue because it is considered the flagship software engineering conference \cite{Damian&Zeller:2022} and the proceedings of ICSE are easily accessible online \cite{ICSE2020Proceedings}. We collected our sample from ICSE 2019, 2020 and 2021 technical tracks to enrich the sample of studies and ensure our findings cover multiple editions of the conference. 
In total, the technical track of ICSE 2019, 2020, and 2021 contained 379 papers.
Sampling three venues were also in line with software engineering meta-studies, such as the study of Storey \textit{et al.}~\cite{Storey&Ernst:2020}, which used two venues, ICSE 2017 technical track and ESME 2017 paper, as their sample.

For completeness, we list below all of the inclusion criteria used for the data set of our study:
\begin{itemize}
    \item Technical research track papers from ICSE 2019, 2020 and 2021.
    \item Studies that contained participants such as surveys, control groups, action research, grounded theory, focus groups, interviews, field studies, lab studies, validation studies, task-based studies and judgement studies.
    \item Studies with one or more participants
\end{itemize}

\subsection{Selecting Papers with Participant Studies}
\label{sec:selectingParticipantStudies}

Software engineering research employs a variety of research methods. As we are interested in studies with participants, we need to identify papers that have employed a particular set of methods, e.g., surveys, interviews, field studies, etc. To this aim, we read the abstracts of all 376 papers in the technical research track of ICSE, from 2019 to 2021.

We choose to read the abstracts of each paper in order to find research papers containing participant studies. Abstracts often convey the main methodology of a research therefore if a participant study was conducted as part of a research, it would likely be mentioned in the abstract. This is particularly the case with participant studies, as the inclusion of interviews and surveys with practitioners is well-valued by the SE community. Analyzing paper abstracts is commonly used in related literature when conducting similar content analyses~\cite{Shaw:2003,Perera&Hussain:2020}. Given our initial sample contains 379 papers, reading the abstracts was a suitable and efficient method for filtering papers that are candidates for containing participant studies.  

To select papers with participant studies, we identified words and phrases that indicated that the research involves research participants. Some examples of these words and phrases are given in Table.~\ref{tab:participant-studies-keywords}.

\begin{table}[]
    \centering
    \caption{The table shows some of the words and phrases we recognized in abstracts of papers containing participant studies.}
    \label{tab:participant-studies-keywords}
    \begin{tabularx}{\linewidth}{X}
    \toprule
    \textbf{Keywords indicating participant studies} \\
    \midrule
    judgement study, empirical studies, controlled experiment, control group, participants, analyze recordings, grounded theory methodology, grounded theory study, practitioners, surveys, surveyed, interviewed, collected quantitative information, focus group, action research, qualitative insights, quantitative survey, professional software developers, examined [...] developers, developers real work
    \\
    \bottomrule
\end{tabularx}

\end{table}

Sometimes, the abstract of a paper does not explicitly describe the entire methodology adopted by the studies. In the case of only implicit indication of research involving participants, we read the entire paper for verification. For example, in the following abstract, the indication of a study with research participants is not clear. 
\begin{quotebox}
   \textbf{``Using screencasts of developers' real work}, we demonstrate the usefulness of our technique in a practical application for action-aware extraction of key-code frames in developers' work.''  \cite{Zhao&Xing:2019}
\end{quotebox}
The highlighted text here indicates a possibility that the ``developers'' mentioned in the abstract are participants in this study. In such a case, we go beyond reading the abstract to verifying the contents of the paper to see if participants were actually involved in the study. 

We did not consider studies where participants were not directly or actively involved in the research. For example, Wu \textit{et al.} \cite{Wu&Deng:2021} analyzed app user reviews in their research. Since the authors use already available reviews from the users and do not contact the app users directly to gather data from them, we do not consider this paper to have a participant study. 

\begin{table}[H]
    \centering
    \caption{The table shows the number of abstracts we read to select our sample, the number of research papers in our sample and the number of studies we found per venue.}
    \label{tab:summary-of-sample}
    \begin{tabular}{l c c c c}
    \toprule
    \textbf{} & \textbf{2019}  & \textbf{2020} & \textbf{2021} & \textbf{Total} \\
    \midrule
    Number of Abstracts & 109 & 129 & 138 & 376 \\
    Number of Research Papers & 28 & 25 & 26 & 79\\
    Number of Studies & 34 & 33 & 38 & 105\\
    
    \bottomrule
\end{tabular}

\end{table}

Of a total of 376 papers from the three venues of ICSE 2019, 2020, and 2021 technical tracks, we identified 79 publications that met our selection criteria to be included in our study. A summary of this data is presented in Table~\ref{tab:summary-of-sample}. Out of our sample of 79 research papers with participants, some of the papers included more than one participant study. For example, Miller \textit{et al.} \cite{Miller&Rodeghero:2021} conducted two different surveys with different sets of participants. 
Since the two surveys that were described in the paper had different sets of participants, we coded the two different surveys as two different participant studies, noting that they are part of the same research paper. Similarly, several other research papers consisted of two or more studies with participants. Therefore, even though we selected 79 research papers with participant studies as our sample data, we found a total of 105 studies with participants in those research papers. Thus, our main sample consists of a total of 105 participant studies.

\subsection{Developing the Diversity Coding Scheme}
\label{sec:developingCodingScheme}

To the best of our knowledge, there are no other studies that attempt to characterize diversity awareness of SE research. Hence, we cannot reuse an established framework to conduct our study. We start the study by developing an appropriate framework for encoding different diversity categories by conducting a pilot study using the ICSE 2020 technical research track. 
The ICSE 2020 technical research track contains a total of 129 papers. We read the abstracts of the 129 papers and identified 25 research papers with participant studies. 

Once we selected our sample of 25 research papers, the first author read each research paper and identified references creating diversity categories and developing a coding scheme. By following an inductive approach, codes were developed while reading the papers. We used an open coding method where the codes emerged directly from the content of the paper \cite{Saldana:2009}. In the next step, we used the axial coding method where codes were grouped into diversity categories \cite{Saldana:2009}. Tamtik \cite{Tamtik:2022} and Storey \textit{et al.} \cite{Storey&Ernst:2020} used a similar approach in their study. The coding scheme grew as new diversity categories were identified in the sample data. 

Throughout the pilot study, the research team met weekly between February 2021 and June 2021 to discuss each paper in our sample. We discussed, defined and refined the diversity categories that emerged from each paper. 
The initial coding scheme developed through our pilot study which was agreed upon by the research team. While the pilot study was fundamental for developing the initial coding scheme, the coding scheme was further refined and new codes were added to the coding scheme as we went through the rest of our sample. All new codes added, as the coding book evolved, were discussed and agreed upon by the research team.

\subsection{Qualitative Analysis of Papers to Create Diversity Categories}
\label{sec:creatingDivCategories}

 The developed coding scheme mentioned in section~\ref{sec:developingCodingScheme} was used to code the sample of 105 studies. The first author read each paper and identified different words and phrases that describe and distinguish participants in the studies. For example, in the following quote from Krueger \textit{et al.}: 
\begin{quotebox}
    ``When participants elected to participate in the study, we collected basic demographic data (\textbf{sex}, \textbf{gender}, \textbf{age}, \textbf{cumulative GPA}, and \textbf{years of experience}) and \textbf{socioeconomic status} (SES) data.'' \cite{Krueger&Huang:2020}
\end{quotebox}
the words highlighted in bold are descriptive words since they are used to describe the participants. We marked such descriptive words in each research paper. 

We grouped codes to create the diversity categories. For example the words ``sex'', ''gender'', ``gender fluid'', ``male'', ``female'', ``women'', ``sexual orientation'' were all grouped into the diversity category labeled ``gender/sex''. The research team met each week to discuss any potential ambiguities that arouse in forming the diversity categories and find agreement. 

These diversity categories were then added to our coding scheme. Whenever we discovered a descriptive word or phrase that could not previously be included in our coding scheme, we either added the new code to an existing category or added a new category to the coding scheme. Thus, our coding scheme included the diversity categories identified in our sample papers. It is important to note here that the diversity categories in our coding scheme are not all the diversity categories that exist. The ones in our scheme are just the ones we created based on the words and phrases we identified in our sample data. We wanted to inductively create diversity categories to add to our coding scheme rather than using preexisting frameworks for diversity because we did not want to limit our view of diversity to the already existing diversity category definitions. This approach is commonly applied when using the open coding method \cite{Saldana:2009}. 

\subsection{Identify the Function of Diversity Categories in Software Engineering Research}
\label{sec:divCategoriesInSections}

The goal of RQ3 is to understand how diversity is embedded in SE research.
We want to identify the function that participant diversity serves in the selected SE studies.  
We consider four different functions of diversity in participant studies: 1) describing participants, 2) analyzing the impact of diversity in the study results, 3) reflecting upon participant diversity in the study conclusion, and 4) assessing the limitation of diversity in the participant study.   

\begin{figure}[]
 \frame{\includegraphics[scale=0.4]{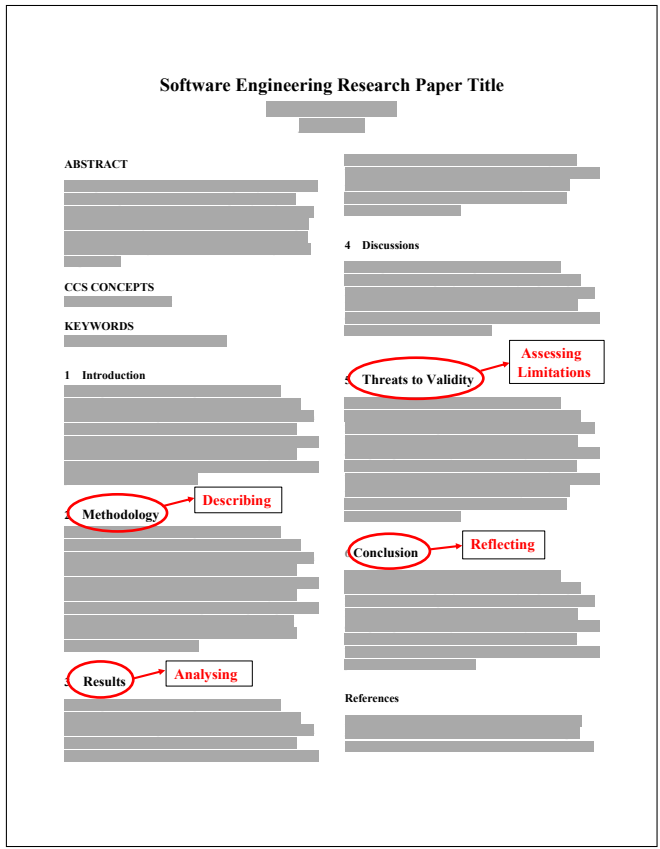}}
    \caption{This image shows a generic layout format of a typical ICSE paper. The sections circled in red are the sections that typically correspond to the four functions of diversity, namely, describing, analyzing, reflecting and assessing limitations in a study.}
    \label{fig:ICSE-format}
\end{figure}

Given our sample contains 105 studies covering a multitude of different study types and goals, we evaluate the function of diversity in a study using a two-step approach. 
First, we identify the section in which diversity is discussed, as the study section also serves a clear function in the study report. 
For example, we illustrate how a hypothetical ICSE paper would have its sections classified by our method in Fig.~\ref{fig:ICSE-format}.  
Second, we carefully analyze the context in which diversity is inserted to confirm its function in the study. 
In detail, we proceed as follows to identify the four functions of diversity:
\begin{itemize}
    \item \textbf{Describing} the diversity of participants. 
    This information is usually reported in the methodology sections of a research paper to describe the diverse sample of participants, e.g., in the ``experimental design'', ``study design'', and ``research design'' sections. 
    In some studies, the diversity information of participants is not reported in the methodology section but the user evaluation part of their study. Hence the importance to consider the context in which diversity is described to identify its function in SE participant studies.  
    
    \item \textbf{Analysing} the diversity of participants in study results. 
    We evaluated the results reported in a research paper to identify if authors analyze the impact of participant diversity when discussing the results of their research. This information is usually reported in the ``results'' section of the paper but can also be found in ``experimental results'', ``findings'', etc. 

    \item \textbf{Reflecting} upon the diversity of participants when concluding the study. 
    We identified if authors reflect on participant diversity and use this data to draw outcomes or conclusions in a paper. 
    This information is typically discussed in the sections named ``discussion'', ``conclusions'' and ``future work''. 

    \item Assessing the \textbf{limitations} of participant diversity in the study. 
    To identify if authors assess the limitations of participant diversity or the lack of collection of diversity data about participants, we evaluate the ``threats to validity'' or ``limitations'' section of research papers. It is important to identify if authors assess the limitations of their participant diversity, as it reflects how aware authors are about diversity or lack thereof.
\end{itemize}

\section{Results}
\label{sec:results}

The goal of our overall research is to assess the diversity awareness in software engineering research with participants. To do so, we first want to know if researchers report any kind of diversity data about participants (RQ1); what kind of diversity they report (RQ2); and the function of diversity in the software research (RQ3). The aforementioned research questions help us assess the awareness of diversity in software engineering participant research. 

\subsection{RQ1. \rqone}

The first step to understanding the awareness of diversity in SE research is to examine how often SE studies report the diversity of their participants. 
This analysis will help us understand how well-established are the practices of collecting, analyzing and reporting demographic data about participants.
Hence, to answer RQ1, we performed the qualitative content analysis described in Section~\ref{sec:creatingDivCategories}. 
We also break our results by type of participant study (e.g., survey, task-based) and for this purpose we rely on how researchers describe their own study. 

\textbf{We found that 98.01\% of the examined participant studies report the diversity of participants.} Out of our sample of 105 studies that we analyzed in our research, 103 studies used descriptive words and phrases when characterizing participants. Only 2 studies out of 105 did not report any participant diversity. 

To make better sense of the results, we break our analysis per type of participant study in Table~\ref{tab:study-categories}. 
First, we note that SE papers use participant studies on 11 different study modalities, the most common being: surveys (51 studies), task-based (33 studies), interviews (22 studies), and validation (19 studies). 
Second, most types of studies tend to report an average of 3 different categories when describing the participants, with Lab studies reporting the highest average (4.6 categories) across the samples we investigated. 
Some studies, such as tasked-based and lab studies, reported a wide array of diversity categories when describing participants, including 8 different types of descriptors.   

Our results indicate that the practice of reporting participant diversity is well-established in participant studies in Software Engineering. 
Researchers tend to select between 2-4 categories of descriptors when describing participants. 
Only in rare cases (2 out of 105 studies) do researchers omit to describe the participants included in their study.     

\begin{table}[]
    \centering
    \caption{The frequency in which studies report diversity category (Reported column) per type of participant study.}
    \label{tab:study-categories}
    \begin{threeparttable}
\begin{tabular}{l|rr|rr}
         \toprule
         \textbf{Study Type} & 
         \multicolumn{2}{c}{\textbf{\# of Studies}}
         & 
         \multicolumn{2}{c}{\textbf{\# Diversity Categories}}
         \\
         & \textbf{Total} & \textbf{Reported} &
         \textbf{Average} & \textbf{Max} \\
        \midrule

         Surveys         & 51 & 49 & 3.0 & 7 \\
         Task based	     & 33 & 32 & 3.3 & 8 \\
         Interviews	     & 22 & 22 & 3.0 & 7 \\
         Validation	     & 19 & 19 & 2.8 & 6 \\
         Lab study	     & 9 & 9 & 4.6 & 8 \\
         Field study     & 6 & 6 & 2.8 & 6 \\ 
         Grounded theory & 4 & 4 & 3.2 & 6 \\
         Control Group   & 2 & 2 & 2.5 & 3 \\
         Action Research & 1 & 1 & 4.0 & 4 \\
         Focus groups    & 1 & 1 & 2.8 & 6 \\
         Judgement study & 1 & 1 & 3.0 & 3 \\

         \bottomrule
         
    \end{tabular}

\end{threeparttable}
\begin{tablenotes}
    The number of studies surpasses 105, as a participant study may involve multiple study types (e.g., interviews and surveys). The study type is based on how the authors describe their own study. 
\end{tablenotes}
\end{table}

\newcommand{\cat}[1]{\textit{#1}}
\newcommand{\terms}[1]{\textbf{#1}}

\subsection{RQ2. What diversity categories are reported in SE participant studies?}

\begin{table*}[ht]
  \centering
    \caption{Diversity categories identified in participant study description of 105 studies from ICSE 2019, 2020, and 2021.} 
    \label{tab:diversity-categories-with-definition}
    \def\profesh#1{\color{RoyalBlue!70}\rule{#1in}{8pt}}
\def\social#1{{\color{RoyalBlue!70}\rule{#1in}{8pt}}}

\ra{1.1}
\begin{tabular}{ c p{0.15\linewidth} p{0.5\linewidth} c l}
    \toprule
    \textbf{Themes} & \textbf{Diversity Category} & \textbf{Definition} & \textbf{Freq.} & 
    \\
    \midrule
    
 \multirow{9}{*}{\textbf{Professional}} & \textbf{Experience}            & The reported professional working experience of a participant such as their experience in programming languages, industrial work experience, number of years of experience in their field.  & 89\%  &  \profesh{0.89}                                      \\

  & \cellcolor{gray!10} \textbf{Main Occupation}       
  & \cellcolor{gray!10} The current occupation of the participant, ex. developer. The current occupation also includes being in graduate school, for example, if the participant's main occupation is currently pursuing their PhD.  
  & \cellcolor{gray!10} 77\% 
  & \cellcolor{gray!10} \profesh{0.77}      \\ 
  
  & \textbf{Education}   & The reported education of a participant could include their level of education, their education institution or their field of study.  & 37\% &  \profesh{0.37}   \\ 
  
\midrule
  
\multirow{12}{*}{\textbf{Social}} 
    & \cellcolor{gray!10} \textbf{Gender/Sex}                 
    & \cellcolor{gray!10} The reported gender or sex of participants. 
    & \cellcolor{gray!10} 39\%  
    & \cellcolor{gray!10} \social{0.39}   \\ 

  & \textbf{Location}                   & The reported geographical location of participants.   & 33\%    &  \social{0.33}      \\ 
  
 & \cellcolor{gray!10} \textbf{Age}                        
 & \cellcolor{gray!10} The reported age of participants.     
 & \cellcolor{gray!10} 18\%            
 & \cellcolor{gray!10} \social{0.18}         \\ 
 
 & \textbf{Language}                   & The reported spoken or written language of participants.   & 8\%  &  \social{0.08}      \\ 
 
 & \cellcolor{gray!10} \textbf{Nationality}                
 & \cellcolor{gray!10} The reported nationality of participants.    
 & \cellcolor{gray!10} 5\%        
 & \cellcolor{gray!10} \social{0.05}            \\
 
 & \textbf{Race}                       & The reported race of participants.       & 4\%          &  \social{0.04}            \\ 
 
 & \cellcolor{gray!10} \textbf{Physiological}   
 & \cellcolor{gray!10} The reported physiological characteristics such as right-handedness or physical disability such as colour-blindness.    
 & \cellcolor{gray!10} 2\% 
 & \cellcolor{gray!10}  \social{0.02} \\
 
 & \textbf{Psychological}       
 & The reported psychological measurements and neurological data of participants such as Optimism, Pessimism, Extroversion, Conscientiousness, Neuroticism, etc.   & 2\% 
 &  \social{0.02}   \\ 
 
 & \cellcolor{gray!10} \textbf{Socioeconomic status}       
 & \cellcolor{gray!10} The reported socioeconomic status of the participant.   
 & \cellcolor{gray!10} 1\%   
 & \cellcolor{gray!10} \social{0.01}   \\ 
    
    \bottomrule
\end{tabular}

\end{table*}

After determining that most SE participant studies do report aspects of participant diversity, we want to identify what kinds of diversity are reported. Such analysis will help us better understand what characteristics of participant diversity are emphasized by SE researchers and the aspects that might be overlooked and require more attention.
To answer RQ2, we created a coding scheme where we identified descriptors about participants in SE studies. We then grouped similar descriptors to form diversity categories as mentioned in section \ref{sec:creatingDivCategories}.

\textbf{We found 12 different diversity categories reported in SE participant studies.}
Table~\ref{tab:diversity-categories-with-definition} shows the diversity categories we identify in the content analysis, as well as their definition and frequency (Freq) in our sample. The frequencies do not sum up to 100\% as, on average, studies report multiple categories of diversities (as shown in Table~\ref{tab:study-categories}).

Overall, two types of diversity remained dominant in the participant diversity reported by our sample of studies, the \cat{Experience} and the  \cat{Main Occupation} of participants. 
This shows that SE studies are often concerned with showcasing a diverse sample of participants regarding professional backgrounds.
The \cat{Experience} of participants was reported in 89\% of the studies. 
We group in this category different kinds of professional experiences, from years of industrial experience to experience in a specific technology, such as programming languages.
As Zhang and Yang report in their study~\cite{Zhang&Yang:2019}:

\begin{quotebox}
    ``Eleven participants had two to \terms{five years of Java experience}, while the other five were \terms{novice programmers with one-year Java experience}, showing a good mix of different levels of Java programming experience.''
\end{quotebox}

\cat{Main Occupation} was reported in 77\% of the studies, showing a well-established practice to collect and report participation occupations. 
This category includes professional occupations (e.g., developer, quality analyst) and education-related occupations (e.g., Ph.D. student, MSc. student).
For example, as Ju \textit{et al.}~\cite{Ju&Sajnani:2021} describes: 

\begin{quotebox}
    ``This process yielded 397 \terms{developers} and 1167 \terms{developer managers} for interviews and 1629 \terms{developers} and 754 \terms{developer managers} for surveys''
\end{quotebox}

Three other categories were reported to describe participants in a comparable frequency: \cat{Gender/Sex} (39\%), \cat{Education} (37\%), and \cat{Location} (33\%).
Reported in at least a third of the studies, these categories help qualify the diversity of participants in terms of gender, scholarly level and geographic location. 
For example, Dias \textit{et al.} report the location diversity of their participants as follows:

\begin{quotebox}
    ``Our interviewees are \terms{located in different countries}, such as Brazil, Canada, Czech Republic, USA, Germany, and Portugal.''~\cite{Dias&Meirelles:2021}
\end{quotebox}

\cat{Age} (18\%) and \cat{Language} (8\%) of participants are less commonly reported in our sample of SE studies. 
The age of participants is a characteristic that may be indirectly captured by participants' years of experience. 
However, it is remarkable that the language spoken by participants is seldom reported, given the international audience of SE papers and the importance of effective communication in any type of study with participants. 
For example, Xia \textit{et al.} reported on the efforts of translating the survey content to ensure effective communication with participants:

\begin{quotebox}
     ``To support respondents from China, we translated our survey to \terms{Chinese} before publishing the survey.''~\cite{Xia&Wan:2019}
\end{quotebox}

Finally, some categories were reported only in a handful of SE participant studies. 
These categories include \cat{Nationality} (5\%), \cat{Race} (4\%), \cat{Physiological} (2\%), \cat{Psychological} (2\%) and \cat{Socioeconomic status} (1\%).
These categories describe specific racial and social aspects of participants and seem to only be collected and reported in specific cases. 
For example, Krueger~\textit{et al.}~\cite{Krueger&Huang:2020} ask participants to complete psychological measurement surveys, such as the Positive and Negative Affect Scale (PANAS, emotional health).
Peitek~\textit{et al.}\cite{Peitek&Apel:2021} reported that all participants in their study had a normal or corrected-to-normal vision and were right-handed.

Once we analyze the frequency of categories based on two major themes of categories, professional and social category (Themes column of Table~\ref{tab:diversity-categories-with-definition}), it becomes clear that \textbf{SE studies emphasize professional diversity over social diversity when selecting and reporting participants.}
All professional categories are reported in at least a third of all studies, while only social categories of \cat{Gender} and \cat{Location} reach similar levels of frequency. 
Social diversity categories, also known as personal or identity categories, are considered particularly important for EDI efforts towards social equity \cite{GovCanada:2021}, however, our results show that SE studies do not often report this type of diversity.

\subsection{RQ3.\rqthree}

In the previous RQ we found that SE studies commonly report characteristics of participants in their studies, with emphasis on professional background. 
In this RQ, we aim to assess the function of participant diversity in SE research. 
To that end, we follow the methodology described in Section~\ref{sec:divCategoriesInSections}, where we classify the function of participant diversity in four categories: 1) describing participants, 2) analyzing the impact of diversity in the study results, 3) reflecting upon participant diversity in the study conclusion, and 4) assessing the limitation of diversity in the participant study.

\begin{figure}[]
    \centering
    \includegraphics[scale=0.67]{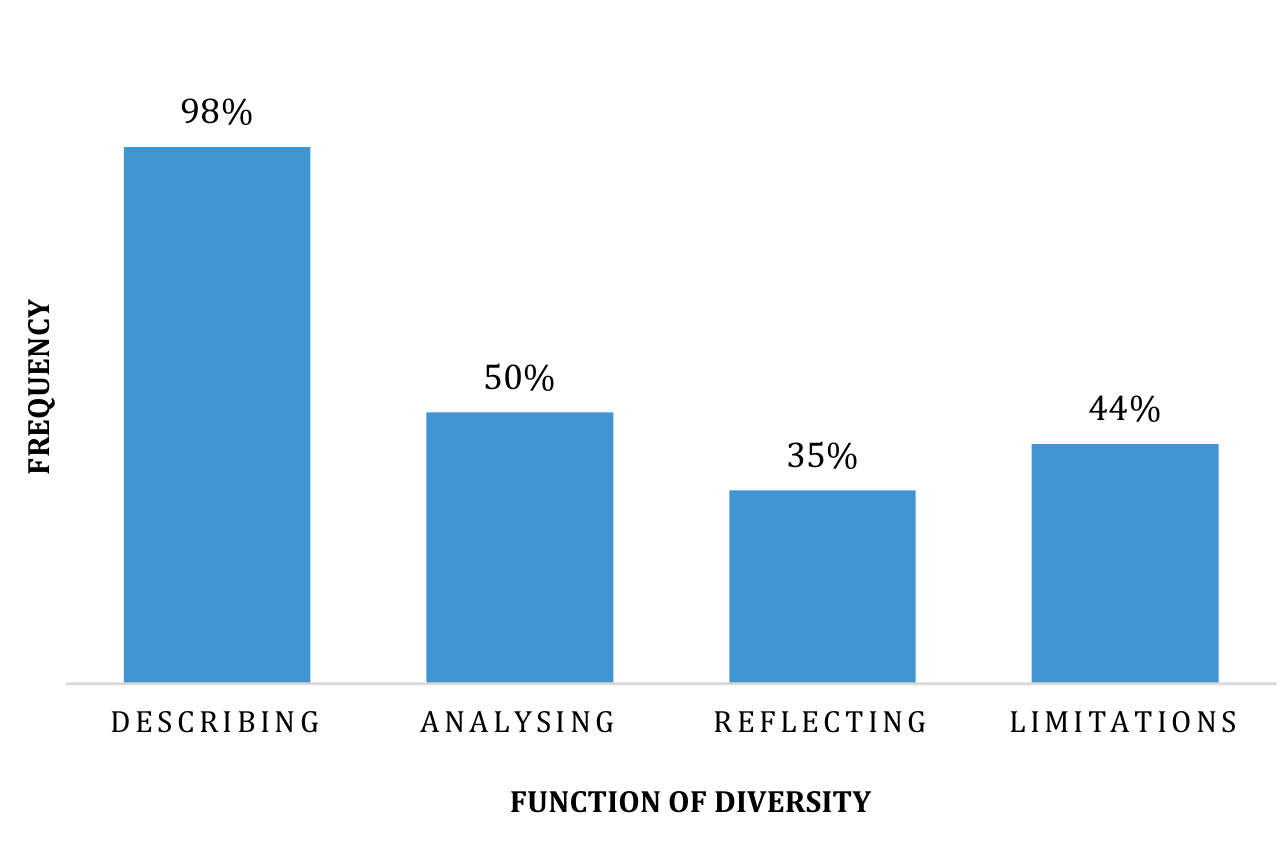}
    \caption{How participant diversity is used in the sample of 105 SE studies. We report the frequency of studies that use participant diversity for \textbf{describing} participants, \textbf{analyzing} results, \textbf{reflecting} the study outcomes, and assessing the study \textbf{limitations}.  }
    \label{fig:diversity-categories-per-section}
\end{figure}

\begin{table}[]
    \centering
    \caption{
    The function of participant diversity in the 105 studies, broken down by diversity category.}
    \label{tab:percent-div-cats-per-section}
    
\newcommand{\first}{\cellcolor{RoyalBlue!45}}
\newcommand{\second}{\cellcolor{RoyalBlue!20}}

\begin{threeparttable}
\begin{tabular}{l c c c c }
    \toprule
    \textbf{Diversity} & 
    \multirow{2}{*}{\textbf{Describing}} & 
    \multirow{2}{*}{\textbf{Analysing}} & 
    \multirow{2}{*}{\textbf{Reflecting}} & 
    \multirow{2}{*}{\textbf{Limitations}} \\

    \textbf{Categories} & & & & \\

    \midrule
    \textbf{Experience}           & \first 82\% & \second 44\% & 28\% & 27\% \\
    \textbf{Occupation}     & \first 76\% & \second 17\% & 10\% & 10\% \\
    \textbf{Gender/Sex}           & \first 37\% & \second 10\% & 6\%  & \second 10\% \\
    \textbf{Education}            & \first 35\% & \second 8\%  & 5\%  & \second 8\%  \\
    \textbf{Location}             & \first 29\% & 9\%  & 4\%  & \second 13\% \\
    \textbf{Age}                  & \first 17\% & \second 4\%  & 2\%  & 2\%  \\
    \textbf{Language}             & \first 5\%  & 1\%  & 0\%  & \second 4\%  \\
    \textbf{Nationality}          & \first 4\%  & \second 1\%  & 0\%  & \second 1\%  \\
    \textbf{Race}                 & \first 4\%  & 0\%  & 0\%  & 0\%  \\
    \textbf{Physiological} & \first 1\%  & \first 1\%  & \first 1\%  & 0\% \\
    \textbf{Psychological}              & \first 2\%  & \second 1\%  & \second 1\%  & 0\% \\
    \textbf{Socioeconomic} & \first 1\%  & 0\%  & 0\%  & 0\%  \\
    \bottomrule
\end{tabular}
\end{threeparttable}

\begin{tablenotes}
    To better visualize the differences across categories, we highlight the most frequent function per category in dark blue and the second most frequent function in a light blue.
\end{tablenotes}
\end{table}

Figure~\ref{fig:diversity-categories-per-section} shows the frequency of every function of participant diversity in the sample of 105 studies. 
We determine that 98\% of studies describe participant diversity by considering 103 out of 105 studies that report diversity. 
Naturally, this result is expected as the primary function of participant descriptors is to describe the participants.
Interestingly, however, participant diversity is less frequently referred to when researchers analyze their study results (50\%), reflect upon the results (35\%), and assess the limitations of their study (44\%).  
This leads us to conclude that \textbf{in most SE participant studies, diversity is reported as a means to describe participants but is less frequently used for further analysis or reflection in the research}.

Table~\ref{tab:percent-div-cats-per-section} displays a breakdown of the participant diversity function, across diversity categories.
To make it easier to identify patterns across the diversity categories, we highlight the highest frequency (dark blue) and the second-highest frequency (light blue) in each category.
It can be seen that \cat{Experience} is the diversity category that most consistently presents the four functions of describing, analyzing, reflecting and assessing limitations in SE studies. 
While also a dominant professional category (as discussed in RQ2), \cat{Main Occupation} is mostly used to describe participants (76\%) and is only used for reflection or assessing limitations in 10\% of the studies in our sample. This leads us to conclude that \textbf{SE researchers use the \cat{Experience} of participants consistently in the four functions of diversity we analyzed}. 
At least a quarter of the studies in our sample discusses the experience of participants when analyzing the results, outcomes, and limitations of their study. 
In contrast, \textbf{all other diversity categories are used mostly to describe participants, with no further reflection on its possible impact in the research.}

Some interesting patterns emerge when we look at the second most frequent function (light blue) per diversity category in Table~\ref{tab:percent-div-cats-per-section}.
We identified \cat{Gender/sex}, \cat{Location}, and \cat{Language} as social diversity categories which researchers use more frequently to assess the study limitations.   
Often, these categories are used to discuss the limitations of the study's generalizability (external validity). 
The following quote from Alsuhaibani \textit{et al.}~\cite{Alsuhaibani&Newman:2021} is an example of the use of participation diversity to discuss the study's limitations:
\begin{quotebox}
    ``With regards to external validity, we did not directly collect geolocation data. However, we did collect IP addresses, which gave us country information. From this we determined that participants came from 72 different countries, mainly from Europe and North America. Thus, we feel that the results are generalizable to a broad population.''~ \cite{Alsuhaibani&Newman:2021}
\end{quotebox}
 
\cat{Race} and \cat{Socioeconomic status} which are two eminent social diversity categories, were only reported in 4\% and 1\% of all studies in our sample, respectively. These two social diversity categories played a role when describing the participants but were not used in further analysis or reflection in any SE study we sampled. This finding leads us to the conclusion that although some SE researchers collect and report data about participants' race and socioeconomic status, they do not use this data for further analysis or reflection in their research. This is not necessarily problematic. For example, considering to which extent the sample of research participants represents the demography is not necessarily linked to investigating social differences. However, especially in the case that participants are far from representing the demography, a further reflection on the impact of this fact on the results of the research might be important.

To conclude, diversity plays four major functions: describing, analyzing, reflecting upon and assessing the limitations of participants in SE studies. Our results indicate that in most studies, SE researchers report diversity when describing participants. However, participant diversity is used less frequently by researchers when discussing the study results, reflecting on the study outcomes, or when assessing the study's limitations.

\section{Discussion}
\label{sec:discussion}

\subsection{Diversity Awareness Model for SE Participant Studies}

\begin{table*}
  \centering
    \caption{Diversity Awareness Model for SE participant studies.}
    \label{tab:diversity-awareness-model}
    
\def\checkmark{\tikz\fill[scale=0.4](0,.35) -- (.25,0) -- (1,.7) -- (.25,.15) -- cycle;} 

\newcommand{\tikzxmark}{%
\tikz[scale=0.23] {
    \draw[line width=0.7,line cap=round] (0,0) to [bend left=6] (1,1);
    \draw[line width=0.7,line cap=round] (0.2,0.95) to [bend right=3] (0.8,0.05);
}}

\ra{1.1}
\begin{threeparttable}
\begin{tabular}{l}

    \toprule
    \textbf{Level 0: Non Identifiable} \\
    \midrule
    
    No reporting of the diversity of participants. \\
    No reporting of considerations of diversity for the selection of participants. \\
    No reflection on the lack of diversity (including statements such as “diversity (or gender) was not considered”). \\

    \midrule
    \textbf{Level 1: Beginner's Level} \\
    \midrule
    
    Mentioning some diversity categories, categories are limited. \checkmark \\ 
    No further explanation on the relevance of the categories. \\ 
    No reporting of considerations of diversity for the selection of participants. \\
    No reflection on the lack of diversity and implications for research. \\

    \rowcolor{gray!10} 
    \hspace{1cm} 
    \textbf{Example:} ``20\% of the research participants were women'' (Study includes no further reflection on participants' gender) \\

    \midrule
    \textbf{Level 2: Intermediate Level} \\
    \midrule
    
    Mentioning of diversity categories. \checkmark \\
    Reporting of the consideration of diversity for the selection of participants. \checkmark \\
    Limited to no reflection on the lack of diversity and implications for research. \\
    Limited to no explanation on the relevance of the categories. \\
    
    \rowcolor{gray!10}
    \hspace{1cm}
    \textbf{Example:} ``Although we tried to reach a balanced representation of gender, only 20\% of the research participants were women'' \\

    \midrule
    \textbf{Level 3: Advanced Level} \\
    \midrule
    
    Reporting the consideration of diversity in the selection of participants. \checkmark \\
    Reflecting diversity (or the lack thereof) and implications for research. \checkmark \\
    Explaining the relevance of the categories. \checkmark \\

    \rowcolor{gray!10} 
    \hspace{1cm} 
    \textbf{Example:} ``We paid attention to diversity in recruiting participants. Since our study involves reaction time, vision, and other physical factors, \\ 

    \rowcolor{gray!10} 
    \hspace{1cm} 
    we paid special attention to including participants of diverse ages and diverse physical abilities.'' \\
    
    \midrule
    \textbf{Level 4: High Level} \\
    \midrule
    
    Extensive reporting of the consideration of diversity in the selection of participants. \checkmark 
     \\
    
    Extensive reflection on diversity (or the lack thereof), implications for research, and improvement of research design. \checkmark \\
    Extensive explanation and reflection of the relevance of particular categories for the research topic. \checkmark \\

    \rowcolor{gray!10}
    \hspace{1cm}
    \textbf{Example:} ``We paid attention to diversity in recruiting participants in terms of gender, race, language, age. Since our study involves  \\
    
    \rowcolor{gray!10}
    \hspace{1cm}
    reaction time, vision, and other physical factors, we paid special attention to include participants of a broad diversity of age and diverse   \\

    \rowcolor{gray!10}
    \hspace{1cm}
    physical ability. However, the study design might have benefited from a greater diversity among the researchers.
    All researchers involved in \\

    \rowcolor{gray!10}
    \hspace{1cm}
    this study are  between 27-40 years old, 80\% of the researchers are male, and no researcher has color vision deficiency \\

    \rowcolor{gray!10}
    \hspace{1cm}
    or a physical disability. \\

    \bottomrule

\end{tabular}

\end{threeparttable}

\begin{tablenotes}
Note that diversity awareness can only be evaluated by considering the entire research report. In general, the level of diversity awareness cannot be determined from a single phrase. These example phrases only serve the purpose of illustrating how the levels can be distinguished from each other.
\end{tablenotes}
\end{table*}

While studying the extent to which SE participant studies report on the diversity of participants, we identified five types of studies which were distinguishable in terms of considering diversity in the overall research and publication approach. Based on these five types of studies, we propose a model for diversity awareness as shown in Table~\ref{tab:diversity-awareness-model}.
The model contains five levels of diversity awareness and it can be helpful in assessing the level of diversity awareness of authors of a participant study and supporting SE researchers in reflecting on diversity. 

In Table~\ref{tab:diversity-awareness-model}, we describe short fictional examples for each level. 
We refrain from providing real examples for all levels, as our goal is not to single out past studies but to discuss paths for improving future ones.  
Good examples, however, can help illustrate and guide SE researchers.
As a good example of a study with high-level diversity awareness, we report on the work of Dias \textit{et al.}~\cite{Dias&Meirelles:2021}.  
The authors of this study demonstrate a high level of diversity awareness (level 4), as they consciously ``choose'' to include diversity and report in detail about their choice and the reasons for it. 

\begin{quotebox}
    ``To foster diversity, when inviting the participants, we prioritize women and non-US based maintainers. We took this decision to avoid having too many “Silicon Valley” participants, as they are over-represented amongst OSS maintainers.'' \cite{Dias&Meirelles:2021}
\end{quotebox}

A few studies in our sample even reported the diversity of the researchers involved. The following quote from Gerosa \textit{et al.}~\cite{Gerosa&Wiese:2021} shows how the authors report the diversity of the researchers. This indicates diversity awareness that goes beyond considering the diversity of research participants.
\begin{quotebox}
    ``we formed an international and diverse team of researchers, who are originally from South America (4), Europe (3), and Asia (1) and were working, at the time of this study, in North America (5), Europe (1), South America (1), and Australia (1). Seven researchers work in academia with extensive experience with OSS, and one researcher is a practitioner working in an OSS company.''~\cite{Gerosa&Wiese:2021}
\end{quotebox}

\subsection{Diversity Awareness Guidelines}

It is difficult to determine the level of diversity awareness quantitatively. For example, reporting several diversity categories or stating that the gender diversity of participants was not considered because the research is not about gender does not necessarily mean a high level of diversity awareness.
\textbf{It is the combination of reporting, analyzing, addressing and reflecting on the considerations of diversity that can demonstrate the diversity awareness of authors.} Here, we propose questions that can guide researchers when self-assessing their diversity awareness that they apply on their research. 
\begin{itemize}
    \item Do you consider diversity in your research?
    \item Which diversity categories did you consider when selecting research participants?
    \item How do you make sure that your research participants are diverse (e.g., a demographic representation of society)?
    \item What recruitment efforts did you undertake to reach a demographic representation?
    \item Which groups/diversity categories are over represented?
    \item Which professional diversity categories do you consider (e.g., work experience, ...)?
    \item Which social diversity categories do you consider (e.g., gender, language, nationality, age, ...)?
    \item Which diversity categories do you NOT consider? Why do you not consider them?
    \item Which diversity categories might be relevant in the context of your research?
    \item Are you reporting your diversity considerations in your research publication?
    \item Are the diversity categories considered relevant to your research?
\end{itemize}

\subsection{Impact on Society}

We believe this research to have a significant impact on the design of future SE research, in particular research that includes research participants. Our work both highlights the importance of considering and reporting the diversity of participants as well as provides guiding questions to successfully integrate considerations of diversity and inclusion into the research design. Furthermore, our proposed model of diversity awareness helps researchers to self-assess and review their research to identify lacks which, otherwise, they would have missed. For research that includes participant studies, defining, selecting and recruiting research participants is an evident opportunity to consider and implement diversity and inclusion into research. The better diversity is considered and implemented in the research design the broader will be the acceptance and the benefit of the research and development for society.

\section{Threats to Validity}
\label{sec:threats}

In this section, we recognize the threats to validity of our research and discuss methods we applied in our research design to mitigate these threats.

\noindent\textbf{Internal Validity:}
A threat to the internal validity of our research is the selection of our sample of participant studies. This process is described in Section~\ref{sec:selectingParticipantStudies}. The first author read all the abstracts from our selected venue to identify participant studies for our sample. 
There is a chance that participant studies were not included because the authors have not described their methodology in their abstract. To mitigate this threat of subjectivity, we took two steps. Firstly, whenever ambiguity arose, the first author read the entire paper to confirm (or not) its inclusion in the study.  Secondly, if after verifying the contents of the paper there was still ambiguity, the research team discussed if the study would be considered part of the sample or removed. Given our sample consists of 105 participant studies, it is unlikely that any missing study would have drastically changed our overall results. 

The classification of diversity categories in the coding scheme for RQ1 and RQ2 could also pose a threat to the internal validity of our findings. This is because this classification was done manually and can be regarded as subjective. However, to mitigate this threat, we took the following steps, described in Section~\ref{sec:developingCodingScheme}. First, we conducted a pilot study and met regularly for the duration of this study in order to discuss the classification of each diversity category in our pilot sample. Second, if ever there was ambiguity during coding, we discussed it among the authors to reach a consensus. 

\noindent\textbf{External Validity:}
A threat to external validity is that our selection sample was limited to the ICSE technical tracks of the 2019, 2020, and 2021 editions. 
ICSE is highly regarded as a top-tier conference in the SE research community and even though choosing three years of sample papers is within industry standards (e.g.,~\cite{Storey&Ernst:2020}), it is possible that our results might differ if we had selected a different venue for our sample. This might limit the generalizability of our results. However, we believe that our contributions of creating a coding scheme to identify diversity categories, the model for diversity awareness, and guidelines for diversity awareness would still remain valid. 

\noindent\textbf{Construct Validity:}
Our coding scheme poses a threat to our construct validity. We defined four different functions of diversity, namely describing, analyzing, reflecting and assessing limitations of participant diversity. These functions were based on our interpretation of what we thought were relevant functions for this research. However, we acknowledge that there could be other characterizations of different functions of diversity in participant studies. Our classification of diversity categories also poses a similar threat where the diversity categories we created could be labelled differently. However, due to the lack of classification schemes in SE that suited our research, we felt the need to create our own classification scheme for diversity categories. Our approach to identifying diversity categories in SE participant studies is one of the contributions of our research that we hope to see in future works. 

\section{Conclusion and Future Work}
\label{sec:conclusion}

In this paper, we examine the diversity considerations in SE participant studies. 
We apply content analysis to investigate participant studies from three ICSE technical tracks, from 2019 to 2021. 
Our investigation focused on understanding 1) the extent to which participants are described, 2) what diversity categories are more prominent in SE research, and 3) the function participant diversity serves in SE studies.  

On one hand, our findings shed light on some positive remarks for the SE research community.
Reporting participant diversity is a well-established practice, with studies reporting on multiple characteristics of participants.
On the other hand, our results also point to some gaps/challenges that may need further addressing.
Studies emphasize participants' professional backgrounds over their social backgrounds, which may prevent important reflections that are needed in a research community.
Furthermore, participant diversity is often only reported initially in the studies, to describe participants, but is seldom analyzed or reflected upon, when researchers discuss their study results, outcomes and limitations.  
Our study sheds light on the strengths and gaps of diversity awareness in SE participant studies and provides a model that future studies can use to improve diversity awareness. This research is as a starting point for discussions about diversity awareness in SE participant research.
There is a large potential for future work to assess the impact of the lack of diversity consideration in SE participant research, and it's effect on representation and inclusion. Our model can also be applied to other engineering fields to uncover differences of reporting participant diversity across various engineering disciplines.

\bibliographystyle{IEEEtran}
\bibliography{bibliography}

\end{document}